\documentclass[12pt]{article}

\usepackage[utf8]{inputenc}
\usepackage[T1]{fontenc}
\usepackage[a4paper, margin=1in]{geometry}
\usepackage{amsmath, amssymb}
\usepackage{graphicx}
\usepackage{longtable}
\usepackage{makecell}
\usepackage{booktabs}
\usepackage{array}
\usepackage{setspace}
\usepackage[hidelinks]{hyperref}
\usepackage[
  backend=biber,
  style=authoryear-comp,
  natbib=true,
  maxbibnames=99,
  uniquelist=false,
  doi=true,
  url=true,
  eprint=false,
  isbn=false,
]{biblatex}
\usepackage[none]{hyphenat}

\DeclareFieldFormat{doi}{\mkbibacro{DOI}\addcolon\space\url{https://doi.org/#1}}
\DeclareFieldFormat{url}{\mkbibacro{URL}\addcolon\space\url{#1}}
\DeclareFieldFormat{eprint:arxiv}{\mkbibacro{arXiv}\addcolon\space\url{https://arxiv.org/abs/#1}}
\AtEveryBibitem{%
  \clearfield{urldate}%
  \clearfield{month}%
  \clearfield{day}%
}
\renewbibmacro*{doi+eprint+url}{%
  \iftoggle{bbx:doi}
    {\printfield{doi}}
    {}%
  \newunit\newblock
  \iftoggle{bbx:eprint}
    {\usebibmacro{eprint}}
    {}%
  \newunit\newblock
  \iftoggle{bbx:url}
    {\printfield{url}}
    {}}

\graphicspath{{figures/}}

\newcommand{\Msun}{M$_{\odot}$}
\newcommand{\Mearth}{M$_{\oplus}$}
\newcommand{\Rearth}{R$_{\oplus}$}
\newcommand{\Wkg}{W\,kg$^{-1}$}
\newcommand{\Mdot}{$\dot{M}$}

\title{\textbf{Power Density as a Universal Biosignature:\\
An Application to the Stellivore Hypothesis}}
\author{Cl\'ement Vidal$^{1,2}$ \and Luciano Kadian$^{3}$}
\date{}

\begin{document}

\maketitle

\begin{center}
\begin{tabular}{l}
$^{1}$ Vrije Universiteit Brussel\\
$^{2}$ Blue Marble Space\\
$^{3}$ University at Buffalo\\
\end{tabular}
\end{center}

\small
To appear in: \emph{Phenomenon of Life on Earth and Beyond: Where Biology and Physics Meet}, edited by George E.\ Mikhailovsky and Abir Igamberdiev. Astrobiology Perspectives on Life in the Universe. John Wiley \& Sons, Limited.

\begin{abstract}
\noindent Power density is a central measure underlying a common thread in cosmic evolution. It measures the rate at which free energy transits in a complex system of a given mass. We argue that it can be seen as a universal signature, and thus a strategy of ``following high power density'' can inform both biosignature and technosignature searches. We use known binary star masses and accretion rates to compute power densities of cataclysmic variables, neutron stars and transient black holes, and show that these accreting binaries have high power densities. We compare this result with other accreting systems in astrophysics as controls: young stellar objects (YSOs) and supermassive black holes (SMBHs). We review possible natural explanations, and then explore the possibility that these accreting binary systems might be instances of Type II stellar civilizations on Kardashev's scale, civilizations feeding on stars, or \emph{stellivores}. Besides power density, we present additional living clues, such as their variety, their existence far from thermodynamic equilibrium, their apparent homeostasis, the need to fine-tune models to explain them, the apparent existence of energy reserves, accretion control and ejection control, as well as other potential living systems functions. We summarize this case using living systems theory and discuss implications for big history.

\vspace{1em}
\noindent\textbf{Keywords:} Astrobiology, agnostic biosignatures, technosignatures, high-energy astrophysics, power density, stellivores, binary stars, cataclysmic variables, neutron stars, black holes.

\vspace{1em}
\noindent Correspondence $|$ Cl\'ement Vidal, \href{mailto:contact@clemvidal.com}{contact@clemvidal.com}
\end{abstract}

\begin{quote}
\emph{Biological complexification leads to} high energy, far-from-equilibrium systems\emph{, rather than the lower energy, equilibrium systems that are the target of non-biological complexification, so in that fundamental sense the two are quite distinct.}\\[0.5em]
\null\hfill\citep[pp.~153--154]{pross2005EmergenceBiological}
\end{quote}

\section*{Introduction}

The concepts of energy, energy flows, and thermodynamics have proven universally applicable to study astrophysical, biological, sociological or technological systems. In particular, the rise of \emph{power density} through cosmic evolution is the central unifying theme of big history \citep[see e.g.][]{spier1996StructureBig,chaisson2001CosmicEvolution,christian2004MapsTime,judson2017EnergyExpansions}. Power density measures the amount of energy that flows through a system, normalized by its mass, and its units are watts per kilogram (\Wkg). It is also called \emph{Energy Rate Density} in astrophysics and cosmology, or \emph{mass-specific power} in engineering, and can also refer to power per unit volume (W\,m$^{-3}$) or power per unit area (W\,m$^{-2}$). In this paper, we use the mass-specific power definition of power density (\Wkg).

To have a high power density, a system must not only be able to sustain a high energy flow, but it must also be of a relatively low-mass while surviving this high energy flow. For example, although the Sun has a high energy outflow, it is also very massive, and thus its power density value is low, at about $2 \times 10^{-4}$~\Wkg.

Let us compute the power density of planet Earth as seen from space, as if it were observed by a distant observer. Let us assume that the Earth intercepts sunlight over its disk, i.e. a cross-section $\pi R^2$, with a solar constant $S = 1361$~W\,m$^{-2}$ and \Rearth{} $= 6371$~km; then the intercepted power is:
\begin{equation*}
P = S \times \pi R_{\oplus}^2 \approx 1361 \times 1.27 \times 10^{14} \approx 1.74 \times 10^{17}~\mathrm{W}.
\end{equation*}

Dividing by Earth's mass \Mearth{} $= 5.972 \times 10^{24}$~kg, we have a power density of $2.9 \times 10^{-8}$~\Wkg, four orders of magnitude less than the Sun. Obviously, the biosphere's mass is tiny in comparison to the whole mass of the Earth, which means that computing a finer biospheric boundary would yield higher values.

A complex system such as a human body has a value of about 2~\Wkg, and modern society 50~\Wkg{} \citep[p.~139]{chaisson2001CosmicEvolution}.

Power density is literally universally applicable, because it is a combined measure of matter and energy through time, so it can compare systems as diverse as galaxies, stars, planets, plants, animals, societies or technologies. \citet{chaisson2001CosmicEvolution,chaisson2003UnifyingConcept,chaisson2011EnergyRate} has argued at length that it can be used as a complexity metric, and that power density has consistently risen in cosmic evolution.

A natural application from this universality is that power density also doubles as an agnostic biosignature that would apply to any  extraterrestrial lifeform or technology we might find. This leads to a straightforward agnostic search strategy: \emph{follow high power density systems in the universe.} \citet{chaisson2003UnifyingConcept} has argued that power density is a ``unifying concept for astrobiology'', but it has not yet been applied explicitly in astrobiology with concrete data. What are the known systems with high power densities in the universe? We compute in this paper the power density of some binary star systems in accretion, and show that they score highly.

We review possible explanations, including potential issues with the metric and the data, and use similar accreting systems as controls, such as young stellar objects and supermassive black holes.

Finally, we discuss the hypothesis that such a high power density is consistent with accreting binaries being actually complex, living, advanced civilizations. If so, they would constitute Type II civilizations on Kardashev's \citeyearpar{kardashev1964TransmissionInformation} scale that are feeding on stars, or \emph{stellivores} \citetext{\citealp[chap.~9]{vidal2014BeginningEnd}; \citealp{vidal2016StellivoreExtraterrestrials}; \citealp{haqq-misra2025ProjectionsEarths}}.

\section{Power density of accreting binaries}

We compute here the power density of 130 white dwarfs (WDs), 30 neutron stars (NSs), and 19 black holes (BHs). The flow of energy is captured through the \emph{accretion rate} (noted \Mdot), expressed in solar mass per year \Msun.y$^{-1}$ in astronomy (1~\Msun.y$^{-1}$ is equivalent to $6.30 \times 10^{22}$~kg\,s$^{-1}$ in SI units). To obtain a power density, we must divide it by the mass of the primary WD, NS or BH (unit in kilograms or solar mass, \Msun). In astrophysics and astronomy, the accretion rate is not a variable that is directly observed, but is derived from an accretion model. The mass determination of the compact star (WD, NS, or BH) is also subject to uncertainties.

To compute power density, we also face the issue of determining how much of the accreted matter is converted into energy. A 100\% conversion efficiency would mean perfect matter-to-energy conversion. As a first approximation, we can use the property that efficiency of energy conversion simply depends upon how compact the star is, via the formula:
\begin{equation}
\eta = \frac{GM}{c^2 R}
\end{equation}
where $G$ is the gravitational constant, $c$ the speed of light, $M$ and $R$ the mass and radius of the compact star. Given that WDs in our data selection weigh on average 0.78~\Msun, assuming a 5000~km radius, this leads to an efficiency $\eta = 2.3 \times 10^{-4}$. This results in an average power density of $4.79 \times 10^{-3}$~\Wkg. However, one can argue that accretion onto WDs is not the primary source of energy and that one should take instead nuclear burning as the main source of energy \citep[p.~2]{frank2002AccretionPower}. In that case, nuclear fusion has an efficiency of $\eta = 0.007$ and the power density in our sample becomes much higher, 0.144~\Wkg{} (see Table~\ref{tab:wd}). Note also that our dataset \citep{dubus2018TestingDisk} excludes polars, which are accreting white dwarfs that accrete directly via their magnetic field lines instead of through an accretion disk. This configuration does not allow the determination of the accretion rate based on the disk luminosity.

\begingroup
\scriptsize
\setlength{\tabcolsep}{4pt}
\renewcommand{\arraystretch}{0.95}
\begin{longtable}{@{}llll@{}}
\label{tab:wd}\\
\toprule
\textbf{Name} & \textbf{Mass (kg)} & \textbf{Accretion rate (kg\,s$^{-1}$)} & \textbf{Power density (W\,kg$^{-1}$)} \\
\midrule
\endfirsthead
\toprule
\textbf{Name} & \textbf{Mass (kg)} & \textbf{Accretion rate (kg\,s$^{-1}$)} & \textbf{Power density (W\,kg$^{-1}$)} \\
\midrule
\endhead
\bottomrule
\endfoot
\bottomrule
\addlinespace[4pt]
\caption{Power density of 130 white dwarfs in accretion. Systems from AE~Aqr downwards are nova-like; those above are cataclysmic variables. Note that we have not reported in this table the error bars in the mass and accretion rate ($\dot{M}$) estimates, but we plotted them in Fig~\ref{fig:powerdensity}. We assumed an accretion efficiency of $\eta = 0.007$. Data from \citet{dubus2018TestingDisk}.}\label{tab:wd}
\endlastfoot
2325+8205 & $1.49 \times 10^{30}$ & $1.05 \times 10^{14}$ & 0.0443 \\
AB Dra & $1.59 \times 10^{30}$ & $1.10 \times 10^{14}$ & 0.0435 \\
AH Her & $1.89 \times 10^{30}$ & $1.70 \times 10^{14}$ & 0.0566 \\
AM Cas & $1.09 \times 10^{30}$ & $9.50 \times 10^{13}$ & 0.0546 \\
AQ Eri & $1.29 \times 10^{30}$ & $1.40 \times 10^{13}$ & 0.006815 \\
AR And & $1.49 \times 10^{30}$ & $1.00 \times 10^{14}$ & 0.0422 \\
AT Ara & $1.05 \times 10^{30}$ & $1.60 \times 10^{15}$ & 0.9551 \\
AT Cnc & $1.49 \times 10^{30}$ & $2.00 \times 10^{14}$ & 0.0844 \\
AY Lyr & $1.49 \times 10^{30}$ & $4.00 \times 10^{13}$ & 0.0169 \\
BD Pav & $2.29 \times 10^{30}$ & $7.00 \times 10^{12}$ & 0.001926 \\
BF Ara & $1.33 \times 10^{30}$ & $7.50 \times 10^{13}$ & 0.0354 \\
BF Eri & $2.55 \times 10^{30}$ & $1.60 \times 10^{13}$ & 0.003955 \\
BI Ori & $1.49 \times 10^{30}$ & $5.00 \times 10^{13}$ & 0.0211 \\
BV Cen & $2.47 \times 10^{30}$ & $1.35 \times 10^{14}$ & 0.0345 \\
BV Pup & $1.49 \times 10^{30}$ & $1.75 \times 10^{14}$ & 0.0738 \\
BX Pup & $1.49 \times 10^{30}$ & $9.00 \times 10^{13}$ & 0.038 \\
BZ UMa & $1.51 \times 10^{30}$ & $3.60 \times 10^{12}$ & 0.001499 \\
CH UMa & $2.51 \times 10^{30}$ & $1.00 \times 10^{13}$ & 0.002511 \\
CN Ori & $1.47 \times 10^{30}$ & $1.00 \times 10^{15}$ & 0.4276 \\
CW Mon & $1.49 \times 10^{30}$ & $4.00 \times 10^{13}$ & 0.0169 \\
CY Lyr & $1.49 \times 10^{30}$ & $1.70 \times 10^{14}$ & 0.0717 \\
CY UMa & $1.37 \times 10^{30}$ & $1.05 \times 10^{13}$ & 0.004815 \\
CZ Ori & $1.09 \times 10^{30}$ & $2.30 \times 10^{14}$ & 0.1323 \\
DI UMa & $1.49 \times 10^{30}$ & $2.20 \times 10^{13}$ & 0.009281 \\
DO Dra & $1.65 \times 10^{30}$ & $2.10 \times 10^{12}$ & 0.0008005 \\
DX And & $1.49 \times 10^{30}$ & $1.30 \times 10^{14}$ & 0.0548 \\
EI Psc & $1.29 \times 10^{30}$ & $5.00 \times 10^{11}$ & 0.0002434 \\
EM Cyg & $1.99 \times 10^{30}$ & $1.05 \times 10^{15}$ & 0.3322 \\
ER UMa & $1.45 \times 10^{30}$ & $9.00 \times 10^{13}$ & 0.039 \\
ES Dra & $1.15 \times 10^{30}$ & $5.00 \times 10^{13}$ & 0.0273 \\
EY Cyg & $2.19 \times 10^{30}$ & $1.65 \times 10^{13}$ & 0.004746 \\
EZ Lyn & $1.49 \times 10^{30}$ & $2.50 \times 10^{11}$ & 0.0001055 \\
FH Lyn & $1.49 \times 10^{30}$ & $1.10 \times 10^{13}$ & 0.00464 \\
FO And & $1.49 \times 10^{30}$ & $3.30 \times 10^{13}$ & 0.0139 \\
FO Aql & $1.09 \times 10^{30}$ & $4.00 \times 10^{13}$ & 0.023 \\
FO Per & $7.95 \times 10^{29}$ & $3.60 \times 10^{14}$ & 0.2847 \\
FS Aur & $1.09 \times 10^{30}$ & $2.20 \times 10^{13}$ & 0.0127 \\
FT Cam & $1.49 \times 10^{30}$ & $1.25 \times 10^{11}$ & $5.27 \times 10^{-5}$ \\
GD 552 & $1.53 \times 10^{30}$ & $7.00 \times 10^{10}$ & $2.88 \times 10^{-5}$ \\
GY Cnc & $1.97 \times 10^{30}$ & $1.30 \times 10^{13}$ & 0.004155 \\
HL CMa & $1.65 \times 10^{30}$ & $1.75 \times 10^{14}$ & 0.0667 \\
HM Leo & $1.49 \times 10^{30}$ & $2.85 \times 10^{12}$ & 0.001202 \\
HS Vir & $1.35 \times 10^{30}$ & $2.90 \times 10^{12}$ & 0.001349 \\
HW Boo & $1.49 \times 10^{30}$ & $1.15 \times 10^{12}$ & 0.0004851 \\
HX Peg & $1.49 \times 10^{30}$ & $1.50 \times 10^{14}$ & 0.0633 \\
IR Gem & $1.37 \times 10^{30}$ & $7.50 \times 10^{12}$ & 0.003439 \\
IX Dra & $2.78 \times 10^{30}$ & $1.10 \times 10^{13}$ & 0.002486 \\
KS UMa & $1.87 \times 10^{30}$ & $2.25 \times 10^{12}$ & 0.0007573 \\
KT Per & $1.49 \times 10^{30}$ & $9.50 \times 10^{13}$ & 0.0401 \\
KV Dra & $1.41 \times 10^{30}$ & $3.00 \times 10^{12}$ & 0.001337 \\
LL Lyr & $1.49 \times 10^{30}$ & $9.00 \times 10^{13}$ & 0.038 \\
LX And & $1.49 \times 10^{30}$ & $8.50 \times 10^{13}$ & 0.0359 \\
NY Ser & $1.61 \times 10^{30}$ & $3.35 \times 10^{13}$ & 0.0131 \\
OU Vir & $1.39 \times 10^{30}$ & $1.40 \times 10^{13}$ & 0.006328 \\
PU CMa & $9.15 \times 10^{29}$ & $2.45 \times 10^{12}$ & 0.001685 \\
PY Per & $1.49 \times 10^{30}$ & $1.25 \times 10^{13}$ & 0.005273 \\
QW Ser & $1.69 \times 10^{30}$ & $2.65 \times 10^{13}$ & 0.009864 \\
QZ Ser & $1.49 \times 10^{30}$ & $6.00 \times 10^{12}$ & 0.002531 \\
QZ Vir & $1.41 \times 10^{30}$ & $5.50 \times 10^{13}$ & 0.0245 \\
RU LMi & $1.49 \times 10^{30}$ & $2.00 \times 10^{13}$ & 0.008437 \\
RU Peg & $2.11 \times 10^{30}$ & $1.20 \times 10^{14}$ & 0.0358 \\
RX And & $2.27 \times 10^{30}$ & $6.50 \times 10^{13}$ & 0.018 \\
RY Ser & $1.49 \times 10^{30}$ & $5.00 \times 10^{13}$ & 0.0211 \\
RZ LMi & $1.99 \times 10^{30}$ & $2.05 \times 10^{13}$ & 0.006486 \\
SS Aur & $2.15 \times 10^{30}$ & $5.00 \times 10^{13}$ & 0.0146 \\
SS Cyg & $1.61 \times 10^{30}$ & $2.20 \times 10^{14}$ & 0.0859 \\
SS UMi & $1.31 \times 10^{30}$ & $1.20 \times 10^{13}$ & 0.005753 \\
ST Cha & $1.49 \times 10^{30}$ & $7.00 \times 10^{14}$ & 0.2953 \\
SU UMa & $1.35 \times 10^{30}$ & $2.85 \times 10^{13}$ & 0.0133 \\
SV CMi & $1.49 \times 10^{30}$ & $7.50 \times 10^{13}$ & 0.0316 \\
SW UMa & $1.41 \times 10^{30}$ & $9.00 \times 10^{12}$ & 0.004011 \\
SX LMi & $1.33 \times 10^{30}$ & $3.20 \times 10^{12}$ & 0.001511 \\
SY Cnc & $1.49 \times 10^{30}$ & $1.50 \times 10^{15}$ & 0.6328 \\
TT Crt & $1.99 \times 10^{30}$ & $2.05 \times 10^{13}$ & 0.006486 \\
TW Tri & $1.49 \times 10^{30}$ & $1.15 \times 10^{14}$ & 0.0485 \\
TW Vir & $1.81 \times 10^{30}$ & $3.50 \times 10^{13}$ & 0.0122 \\
TY PsA & $1.73 \times 10^{30}$ & $1.50 \times 10^{13}$ & 0.005455 \\
TY Psc & $1.39 \times 10^{30}$ & $4.00 \times 10^{13}$ & 0.0181 \\
TZ Per & $1.49 \times 10^{30}$ & $7.50 \times 10^{14}$ & 0.3164 \\
U Gem & $2.33 \times 10^{30}$ & $4.50 \times 10^{13}$ & 0.0122 \\
UY Pup & $1.49 \times 10^{30}$ & $1.15 \times 10^{14}$ & 0.0485 \\
V1159 Ori & $1.43 \times 10^{30}$ & $8.50 \times 10^{13}$ & 0.0374 \\
V1316 Cyg & $1.43 \times 10^{30}$ & $2.60 \times 10^{14}$ & 0.1143 \\
V1454 Cyg & $1.49 \times 10^{30}$ & $5.50 \times 10^{10}$ & $2.32 \times 10^{-5}$ \\
V1504 Cyg & $1.33 \times 10^{30}$ & $8.50 \times 10^{12}$ & 0.004014 \\
V342 Cam & $1.49 \times 10^{30}$ & $1.20 \times 10^{12}$ & 0.0005062 \\
V355 UMa & $1.49 \times 10^{30}$ & $5.50 \times 10^{12}$ & 0.00232 \\
V392 Hya & $1.49 \times 10^{30}$ & $1.30 \times 10^{14}$ & 0.0548 \\
V426 Oph & $1.79 \times 10^{30}$ & $5.50 \times 10^{13}$ & 0.0193 \\
V478 Her & $1.49 \times 10^{30}$ & $2.00 \times 10^{14}$ & 0.0844 \\
V503 Cyg & $1.45 \times 10^{30}$ & $3.35 \times 10^{13}$ & 0.0145 \\
V513 Peg & $1.49 \times 10^{30}$ & $2.30 \times 10^{13}$ & 0.009703 \\
V516 Cyg & $1.49 \times 10^{30}$ & $5.00 \times 10^{14}$ & 0.2109 \\
V521 Peg & $1.49 \times 10^{30}$ & $3.50 \times 10^{12}$ & 0.001476 \\
V537 Peg & $1.49 \times 10^{30}$ & $1.05 \times 10^{14}$ & 0.0443 \\
V630 Cas & $2.01 \times 10^{30}$ & $7.50 \times 10^{15}$ & 2.35 \\
V713 Cep & $1.49 \times 10^{30}$ & $1.30 \times 10^{11}$ & $5.48 \times 10^{-5}$ \\
V729 Sgr & $1.49 \times 10^{30}$ & $2.05 \times 10^{13}$ & 0.008648 \\
V792 Cyg & $1.49 \times 10^{30}$ & $2.55 \times 10^{14}$ & 0.1076 \\
V811 Cyg & $1.49 \times 10^{30}$ & $2.10 \times 10^{13}$ & 0.008859 \\
V844 Her & $9.15 \times 10^{29}$ & $1.05 \times 10^{13}$ & 0.007222 \\
V893 Sco & $1.77 \times 10^{30}$ & $7.50 \times 10^{12}$ & 0.002666 \\
VW Hyi & $1.33 \times 10^{30}$ & $3.10 \times 10^{13}$ & 0.0146 \\
VW Vul & $6.96 \times 10^{29}$ & $1.95 \times 10^{14}$ & 0.1763 \\
VZ Pyx & $1.59 \times 10^{30}$ & $1.50 \times 10^{13}$ & 0.005932 \\
WW Cet & $1.65 \times 10^{30}$ & $4.00 \times 10^{13}$ & 0.0152 \\
WX Hyi & $1.79 \times 10^{30}$ & $1.85 \times 10^{13}$ & 0.006504 \\
X Leo & $2.05 \times 10^{30}$ & $5.50 \times 10^{13}$ & 0.0169 \\
YZ Cnc & $1.63 \times 10^{30}$ & $4.50 \times 10^{13}$ & 0.0174 \\
Z Cam & $1.97 \times 10^{30}$ & $2.40 \times 10^{14}$ & 0.0767 \\
AE Aqr & $1.25 \times 10^{30}$ & $1.10 \times 10^{14}$ & 0.0552 \\
BG CMi & $1.59 \times 10^{30}$ & $2.00 \times 10^{14}$ & 0.0791 \\
BK Lyn & $8.15 \times 10^{29}$ & $6.50 \times 10^{13}$ & 0.0502 \\
FO Aqr & $1.49 \times 10^{30}$ & $8.50 \times 10^{14}$ & 0.3586 \\
IX Vel & $1.63 \times 10^{30}$ & $3.50 \times 10^{14}$ & 0.135 \\
LQ Peg & $1.49 \times 10^{30}$ & $1.45 \times 10^{15}$ & 0.6117 \\
MV Lyr & $1.45 \times 10^{30}$ & $4.50 \times 10^{14}$ & 0.195 \\
RR Pic & $1.89 \times 10^{30}$ & $1.00 \times 10^{16}$ & 3.33 \\
RW Sex & $1.79 \times 10^{30}$ & $7.00 \times 10^{14}$ & 0.2461 \\
RW Tri & $1.09 \times 10^{30}$ & $1.00 \times 10^{15}$ & 0.5753 \\
TT Ari & $1.79 \times 10^{30}$ & $7.00 \times 10^{14}$ & 0.2461 \\
TV Col & $1.49 \times 10^{30}$ & $7.50 \times 10^{14}$ & 0.3164 \\
TX Col & $1.07 \times 10^{30}$ & $1.20 \times 10^{15}$ & 0.7031 \\
UX UMa & $1.79 \times 10^{30}$ & $4.00 \times 10^{14}$ & 0.1406 \\
V1084 Her & $1.49 \times 10^{30}$ & $4.00 \times 10^{14}$ & 0.1687 \\
V1223 Sgr & $1.85 \times 10^{30}$ & $5.50 \times 10^{14}$ & 0.1871 \\
V3885 Sgr & $1.39 \times 10^{30}$ & $1.05 \times 10^{15}$ & 0.4746 \\
V592 Cas & $1.61 \times 10^{30}$ & $9.50 \times 10^{14}$ & 0.3711 \\
V603 Aql & $2.39 \times 10^{30}$ & $2.40 \times 10^{14}$ & 0.0633 \\
V795 Her & $1.37 \times 10^{30}$ & $4.60 \times 10^{15}$ & 2.11 \\
\midrule
\textbf{Average} & $1.56 \times 10^{30}$ & $3.61 \times 10^{14}$ & 0.1444 \\
\end{longtable}
\endgroup

Turning to NSs and BHs, it is generally assumed that $\eta = 0.1$. Assuming a typical NS mass of 1.4~\Msun, we have an average power density of 0.32~\Wkg{} (see Table~\ref{tab:ns}). In the case of BHs, the data is much sparser (there are fewer known systems, and high uncertainties regarding their masses), but this first approximation results in an average power density of 0.41~\Wkg{} (see Table~\ref{tab:bh}). One could argue that the 0.1 efficiency value is not realistic, as it is an estimate for non-rotating BHs. If a BH rotates up to its maximum speed, the efficiency could reach 0.4 \citep{thorne1974DiskAccretionBlack}. Such efficiency would lead to an average power density of 1.65~\Wkg.

\begingroup
\scriptsize
\setlength{\tabcolsep}{4pt}
\renewcommand{\arraystretch}{0.95}
\begin{longtable}{@{}llll@{}}
\label{tab:ns}\\
\toprule
\textbf{Name} & \textbf{Mass (kg)} & \textbf{Accretion rate (kg\,s$^{-1}$)} & \textbf{Power density (W\,kg$^{-1}$)} \\
\midrule
\endfirsthead
\toprule
\textbf{Name} & \textbf{Mass (kg)} & \textbf{Accretion rate (kg\,s$^{-1}$)} & \textbf{Power density (W\,kg$^{-1}$)} \\
\midrule
\endhead
\bottomrule
\endfoot
\bottomrule
\addlinespace[4pt]
\caption{Power density of 30 neutron stars in accretion. To obtain accretion rates ($\dot{M}$), we used the soft color prior to burst (normalized $F_{\mathrm{per}}$) multiplied by $1.30 \times 10^{-8}$ to obtain values in solar masses per year, from column 15 in table 6 of \citet{galloway2008ThermonuclearType}, and then converted into SI units. We assumed the canonical mass for neutron stars, and an efficiency $\eta = 0.1$.}
\endlastfoot
4U 0513-40 & $2.78 \times 10^{30}$ & $1.50 \times 10^{13}$ & 0.0483 \\
EXO 0748-676 & $2.78 \times 10^{30}$ & $6.30 \times 10^{12}$ & 0.0203 \\
4U 0919-54 & $2.78 \times 10^{30}$ & $2.72 \times 10^{12}$ & 0.00878 \\
4U 1608-52 & $2.78 \times 10^{30}$ & $1.50 \times 10^{13}$ & 0.0484 \\
4U 1636-536 & $2.78 \times 10^{30}$ & $3.10 \times 10^{13}$ & 0.1 \\
MXB 1659-298 & $2.78 \times 10^{30}$ & $3.15 \times 10^{13}$ & 0.102 \\
4U 1702-429 & $2.78 \times 10^{30}$ & $1.64 \times 10^{13}$ & 0.053 \\
4U 1705-44 & $2.78 \times 10^{30}$ & $3.56 \times 10^{13}$ & 0.115 \\
XTE J1710-281 & $2.78 \times 10^{30}$ & $6.66 \times 10^{12}$ & 0.0215 \\
4U 1724-307 & $2.78 \times 10^{30}$ & $2.21 \times 10^{13}$ & 0.0713 \\
4U 1728-34 & $2.78 \times 10^{30}$ & $3.50 \times 10^{13}$ & 0.113 \\
KS 1731-260 & $2.78 \times 10^{30}$ & $4.34 \times 10^{13}$ & 0.14 \\
4U 1735-44 & $2.78 \times 10^{30}$ & $9.00 \times 10^{13}$ & 0.29 \\
SAX J1747.0-2853 & $2.78 \times 10^{30}$ & $3.73 \times 10^{13}$ & 0.12 \\
GX 3+1 & $2.78 \times 10^{30}$ & $1.16 \times 10^{14}$ & 0.375 \\
SAX J1748.9-2021 & $2.78 \times 10^{30}$ & $8.83 \times 10^{13}$ & 0.285 \\
EXO 1745-248 & $2.78 \times 10^{30}$ & $4.42 \times 10^{13}$ & 0.143 \\
4U 1746-37 & $2.78 \times 10^{30}$ & $1.28 \times 10^{14}$ & 0.414 \\
SAX J1750.8-2900 & $2.78 \times 10^{30}$ & $3.43 \times 10^{13}$ & 0.111 \\
GRS 1747-312 & $2.78 \times 10^{30}$ & $2.04 \times 10^{13}$ & 0.0656 \\
SAX J1808.4-3658 & $2.78 \times 10^{30}$ & $9.01 \times 10^{12}$ & 0.0291 \\
GX 17+2 & $2.78 \times 10^{30}$ & $9.65 \times 10^{14}$ & 3.11 \\
3A 1820-303 & $2.78 \times 10^{30}$ & $5.03 \times 10^{13}$ & 0.162 \\
XB 1832-330 & $2.78 \times 10^{30}$ & $6.96 \times 10^{12}$ & 0.0224 \\
Ser X-1 & $2.78 \times 10^{30}$ & $1.94 \times 10^{14}$ & 0.626 \\
HETE J1900.1-2455 & $2.78 \times 10^{30}$ & $5.94 \times 10^{12}$ & 0.0191 \\
Aql X-1 & $2.78 \times 10^{30}$ & $2.26 \times 10^{13}$ & 0.0728 \\
4U 1916-053 & $2.78 \times 10^{30}$ & $1.19 \times 10^{13}$ & 0.0385 \\
4U 2129+12 & $2.78 \times 10^{30}$ & $6.06 \times 10^{12}$ & 0.0195 \\
Cyg X-2 & $2.78 \times 10^{30}$ & $8.87 \times 10^{14}$ & 2.86 \\
\midrule
\textbf{Average} & $2.78 \times 10^{30}$ & $9.93 \times 10^{13}$ & 0.3202 \\
\end{longtable}
\endgroup

\begingroup
\scriptsize
\setlength{\tabcolsep}{4pt}
\renewcommand{\arraystretch}{0.95}
\begin{longtable}{@{}llll@{}}

\label{tab:bh}\\
\toprule
\textbf{Name} & \textbf{Mass (kg)} & \textbf{Accretion rate (kg\,s$^{-1}$)} & \textbf{Power density (W\,kg$^{-1}$)} \\
\midrule
\endfirsthead
\toprule
\textbf{Name} & \textbf{Mass (kg)} & \textbf{Accretion rate (kg\,s$^{-1}$)} & \textbf{Power density (W\,kg$^{-1}$)} \\
\midrule
\endhead
\bottomrule
\endfoot
\bottomrule
\addlinespace[4pt]
\caption{Power density of 19 transient black holes. Note that we have not systematically treated the uncertainties about BH masses, but they are significant. When the upper and lower uncertainties were available, we have taken averages in the plausible mass range. We assumed $\eta = 0.1$, and the data is from \citet{coriat2012RevisitingFundamental}.}
\endlastfoot
GRO J0422+32 & $7.89 \times 10^{30}$ & $1.68 \times 10^{12}$ & 0.00191 \\
A0620-00 & $2.19 \times 10^{31}$ & $3.42 \times 10^{12}$ & 0.0014 \\
GRS 1009-45 & $7.75 \times 10^{30}$ & $1.66 \times 10^{13}$ & 0.0192 \\
XTE J1118+480 & $1.70 \times 10^{31}$ & $1.26 \times 10^{13}$ & 0.00667 \\
GS 1124-684 & $1.49 \times 10^{31}$ & $2.10 \times 10^{13}$ & 0.0126 \\
GS 1354-64 & $1.14 \times 10^{31}$ & $1.03 \times 10^{14}$ & 0.0808 \\
4U 1543-47 & $1.87 \times 10^{31}$ & $7.12 \times 10^{13}$ & 0.0341 \\
XTE J1550-564 & $2.09 \times 10^{31}$ & $1.00 \times 10^{14}$ & 0.043 \\
XTE J1650-500 & $5.43 \times 10^{30}$ & $6.87 \times 10^{12}$ & 0.0114 \\
GRO J1655-40 & $1.25 \times 10^{31}$ & $5.50 \times 10^{13}$ & 0.0394 \\
MAXI J1659-352 & $1.15 \times 10^{31}$ & $8.88 \times 10^{12}$ & 0.00692 \\
GX 339-4 & $1.05 \times 10^{31}$ & $8.00 \times 10^{14}$ & 0.681 \\
4U 1705-250 & $9.41 \times 10^{30}$ & $1.84 \times 10^{13}$ & 0.0176 \\
GRS 1915+105 & $2.78 \times 10^{31}$ & $9.77 \times 10^{15}$ & 3.16 \\
GS 2000+25 & $1.91 \times 10^{31}$ & $4.12 \times 10^{12}$ & 0.00194 \\
V404 Cyg & $2.39 \times 10^{31}$ & $5.90 \times 10^{13}$ & 0.0222 \\
Cyg X-1 & $2.94 \times 10^{31}$ & $8.88 \times 10^{14}$ & 0.271 \\
LMC X-1 & $2.17 \times 10^{31}$ & $4.15 \times 10^{15}$ & 1.72 \\
LMC X-3 & $2.21 \times 10^{31}$ & $4.19 \times 10^{15}$ & 1.7 \\
\midrule
\textbf{Average} & $1.65 \times 10^{31}$ & $1.07 \times 10^{15}$ & 0.4122 \\
\end{longtable}
\endgroup

\begin{figure}[htbp]
\centering
\includegraphics[width=\textwidth]{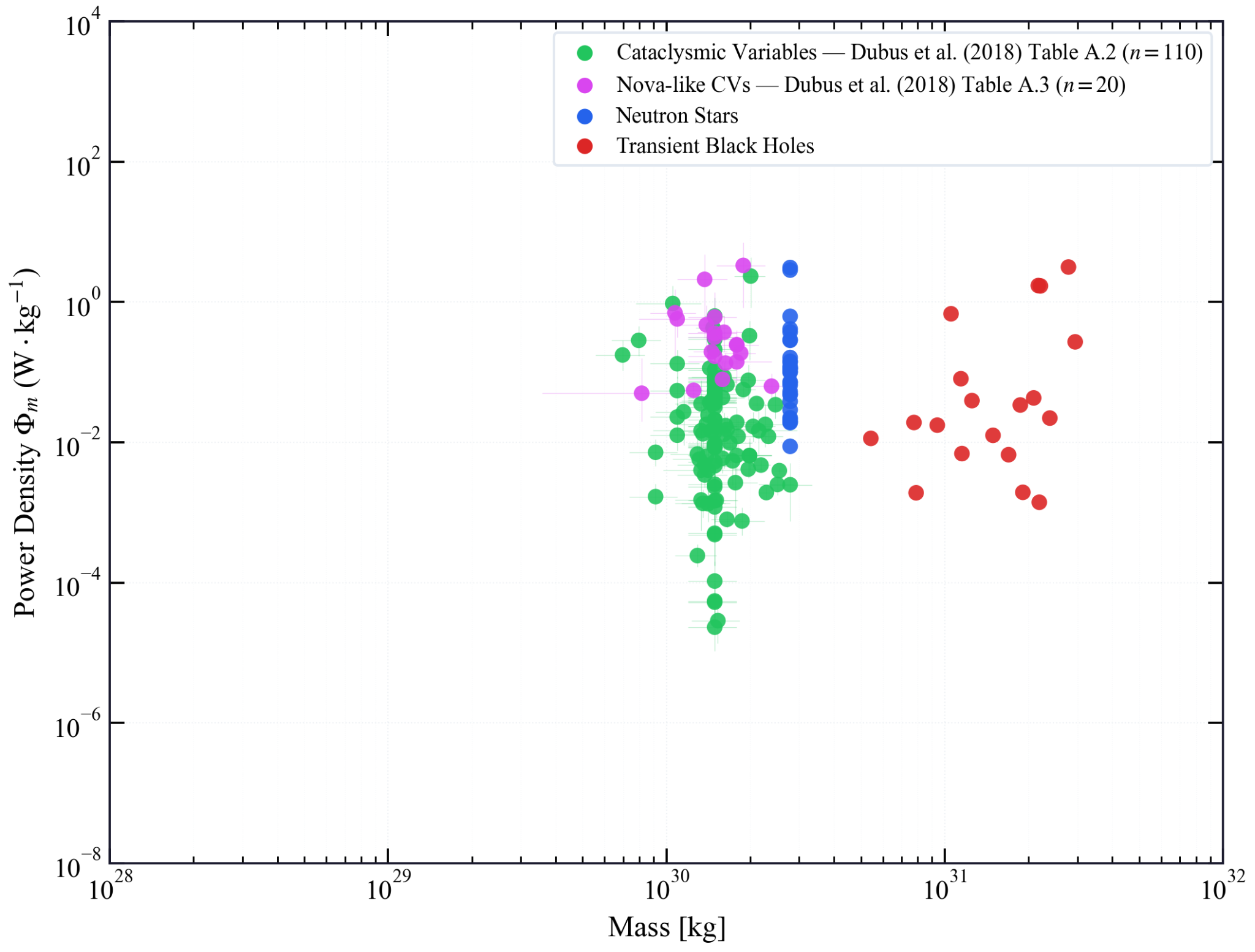}
\caption{Power density for compact accreting binaries: cataclysmic variables (in green) including nova-likes (in pink), but excluding polars; accreting neutron stars (in blue) and transient accreting black holes (in red).}
\label{fig:powerdensity}
\end{figure}

The computations presented here may be refined and thus represent first approximations. But they show high power densities (Fig~\ref{fig:powerdensity}). In the next two sections, we discuss two ways to interpret this result, natural explanations (sec.~\ref{sec:natural}), and an extraterrestrial intelligence interpretation (sec.~\ref{sec:stellivore}).

\section{Possible natural explanations}\label{sec:natural}

\subsection{Control samples}

We can also question whether such high power density is indeed rare or not in astrophysics. For this reason, we computed the power density of two other classes of systems, Young Stellar Objects (YSOs) and Supermassive Black Holes (SMBHs).

\begin{figure}[htbp]
\centering
\includegraphics[width=\textwidth]{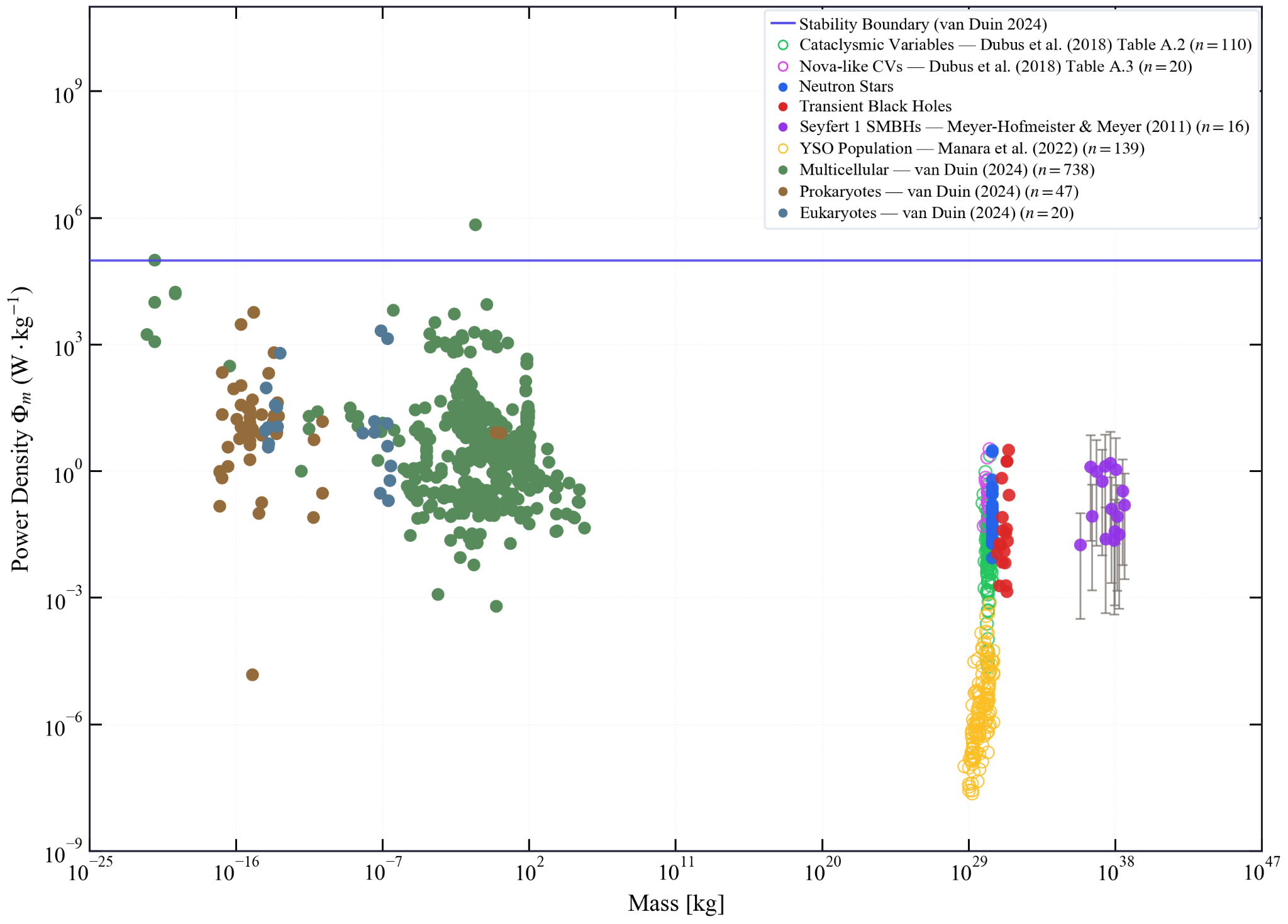}
\caption{Power density values for transient accreting binaries: cataclysmic variables, neutron stars, and black holes, plotted with control datasets, life on Earth on the left of the plot (prokaryotes, eukaryotes, multicellular), young stellar objects (YSOs, yellow), and persistently accreting white dwarfs (nova-like, pink empty circles), and supermassive black holes (magenta filled circles). Above the stability boundary line are explosive systems.}
\label{fig:controls}
\end{figure}

Young Stellar Objects are stars in formation, accreting gas and dust. Their accretion rate is well-studied, and they can thus function as a control sample. We used data from \citet{manara2022DemographicsYoung}, reported in our GitHub repository, and plotted in yellow in Fig~\ref{fig:controls}.

We also computed the power density for 16 supermassive black holes, assuming a non-spinning Schwarzschild black hole, leading to an efficiency of $\eta = 0.057$ (see Table~\ref{tab:smbh}). Note that some models propose values as low as $\eta = 0.001$ \citep{armijo2011GrowthSupermassive}, and the theoretical maximum would be $\eta = 0.42$ for an extreme Kerr black hole. This results in a very high power density.

\begingroup
\scriptsize
\setlength{\tabcolsep}{4pt}
\renewcommand{\arraystretch}{0.95}
\begin{longtable}{@{}llll@{}}
\label{tab:smbh}\\
\toprule
\textbf{Name} & \textbf{Mass (kg)} & \textbf{Accretion rate (kg\,s$^{-1}$)} & \textbf{Power density (W\,kg$^{-1}$)} \\
\midrule
\endfirsthead
\toprule
\textbf{Name} & \textbf{Mass (kg)} & \textbf{Accretion rate (kg\,s$^{-1}$)} & \textbf{Power density (W\,kg$^{-1}$)} \\
\midrule
\endhead
\bottomrule
\endfoot
\bottomrule
\addlinespace[4pt]
\caption{Power density of 16 supermassive black holes (SMBHs) in Seyfert~1 galaxies, data from \citet{meyer-hofmeister2011BroadIron} and references therein.}
\endlastfoot
3C 120 & $1.09 \times 10^{38}$ & $2.31 \times 10^{22}$ & 1.08 \\
NGC 3516 & $8.49 \times 10^{37}$ & $3.78 \times 10^{20}$ & 0.0228 \\
NGC 3783 & $5.87 \times 10^{37}$ & $1.45 \times 10^{21}$ & 0.1266 \\
NGC 4051 & $3.80 \times 10^{36}$ & $6.30 \times 10^{19}$ & 0.085 \\
NGC 4151 & $2.62 \times 10^{37}$ & $1.26 \times 10^{20}$ & 0.0246 \\
Mrk 766 & $6.90 \times 10^{36}$ & $1.32 \times 10^{21}$ & 0.9824 \\
MCG-6-30-15 & $3.08 \times 10^{36}$ & $7.56 \times 10^{20}$ & 1.26 \\
Mrk 590 & $9.52 \times 10^{37}$ & $6.93 \times 10^{20}$ & 0.0373 \\
Mrk 110 & $4.99 \times 10^{37}$ & $1.50 \times 10^{22}$ & 1.54 \\
NGC 4395 & $7.22 \times 10^{35}$ & $2.52 \times 10^{18}$ & 0.0179 \\
NGC 5506 & $1.73 \times 10^{38}$ & $1.07 \times 10^{21}$ & 0.0317 \\
NGC 5548 & $1.34 \times 10^{38}$ & $2.21 \times 10^{21}$ & 0.084 \\
Mrk 509 & $2.88 \times 10^{38}$ & $1.90 \times 10^{22}$ & 0.337 \\
Ark 564 & $1.58 \times 10^{37}$ & $1.76 \times 10^{21}$ & 0.5725 \\
NGC 7469 & $2.45 \times 10^{37}$ & $6.24 \times 10^{21}$ & 1.31 \\
Ark 120 & $3.70 \times 10^{38}$ & $1.13 \times 10^{22}$ & 0.1571 \\
\midrule
\textbf{Average} & $9.03 \times 10^{37}$ & $5.28 \times 10^{21}$ & 0.479 \\
\end{longtable}
\endgroup

To conclude this section, we plot together the power densities of all the systems we have discussed (Fig. \ref{fig:controls}). First, compact binaries (cataclysmic variables, neutron stars, transient black holes). Then, living systems, using the data collected by \citet{vanduin2024DissipativeSystems}. We also include a stability limit identified by van Duin that sits at $10^5$~\Wkg{} as a horizontal line. Above this limit are explosive systems. Turning to the controls, the power density of YSOs is clearly below living systems and accreting binaries, while SMBHs also have a very high power density.

\subsection{Candle objection}

Besides the high power density of SMBHs, one may also object that some trivial systems have high power densities. \citet[p.~144]{chaisson2001CosmicEvolution} discusses candles that indeed have a power density of about 100~\Wkg{} and other high-energy flow situations. This points to a limitation of this metric alone, namely that high power density is only a \emph{necessary but not sufficient} condition to argue that a system is living. In the case of the candle, one should highlight that a candle \textit{is} an artificial system, although of a very simple kind, as there is no \emph{energy control} from the candle itself (it does not switch itself on and off), and the system rapidly reaches equilibrium after the wax is consumed. So it is a dissipative system that tends to and reaches equilibrium instead of actively fighting against thermodynamic equilibrium. Candles obviously also lack other living systems features \citep[sec.~9.5.9]{vidal2014BeginningEnd}.

\section{Stellivore interpretation}\label{sec:stellivore}

In the context of the stellivore hypothesis, we interpret only the primary WD, NS, or BH as living, and the secondary star is just its energy source. In that context it makes sense to normalize only by the mass of the compact object, and not by the mass of the compact object plus its companion star.

A generally agreed upon feature of life is its existence \emph{far from thermodynamic equilibrium}. As Pross writes in the opening quote of this paper, high energy and far-from-equilibrium is the direction of biological complexification.

\begin{figure}[htbp]
\centering
\includegraphics[width=\textwidth]{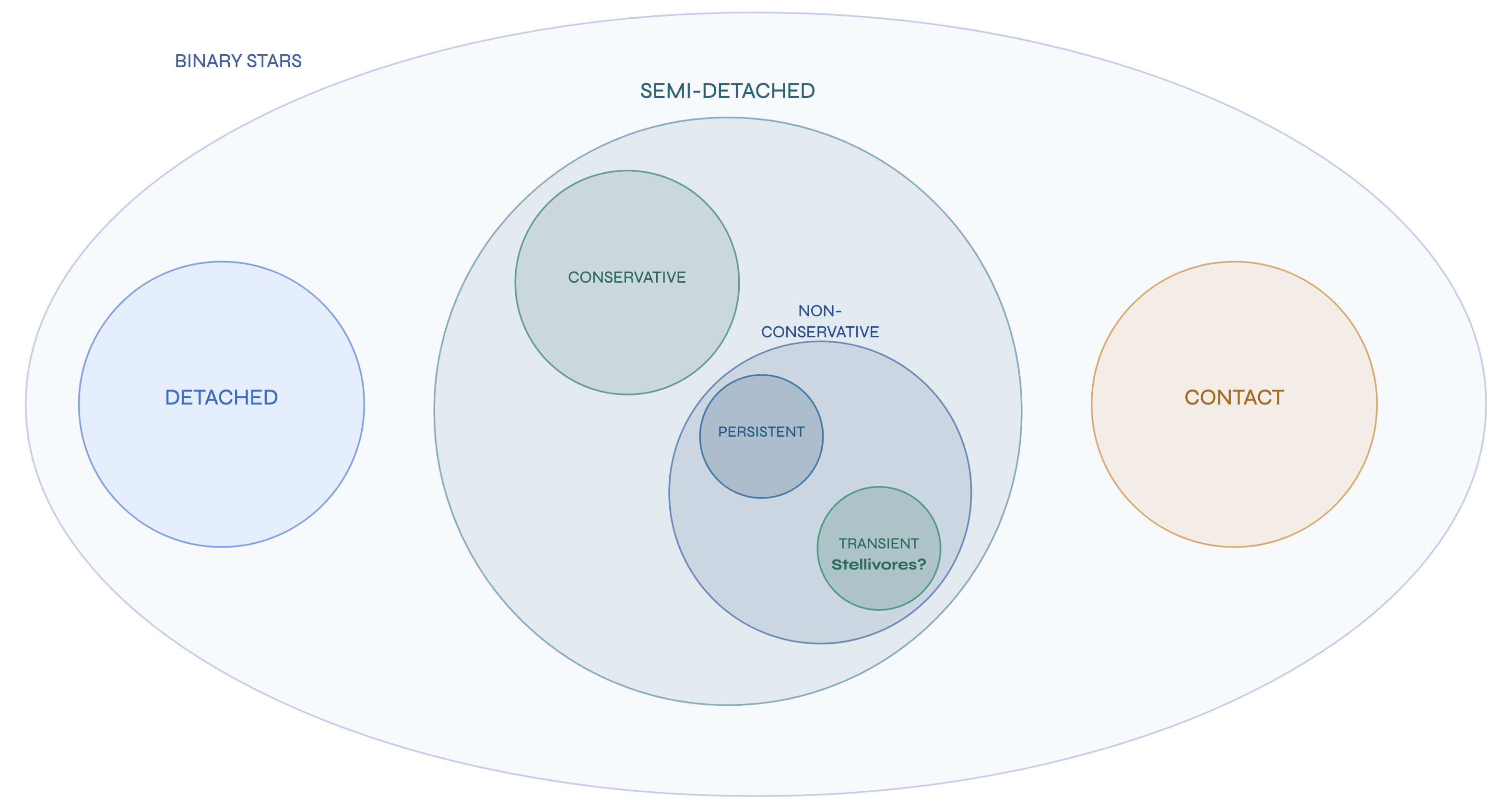}
\caption{Binary stars come in many varieties \citep{kopal1955ClassificationClose}. The binaries that are stellivore candidates are only the ones belonging to the semi-detached, non-conservative, transient category.}
\label{fig:binarytypes}
\end{figure}

In the stellivore interpretation, only a small subset of binary stars are candidate extraterrestrials. \citet{kopal1955ClassificationClose} classifies binaries into three types: \emph{detached binaries}, \emph{semi-detached binaries}, and \emph{contact binaries}. Detached binaries do not interact, and thus have no matter exchanges between them, while contact binaries are destined to merge towards equilibrium, neither of which is a property or a direction of a living system. Only semi-detached binaries remain suspicious, where mass transfer occurs from one star to the other. One should also add the constraint that the systems are open, i.e. that matter is ejected out of the gravitational pull of the binary system. In the astrophysics literature, this is known as a slow non-conservative process \citep[see e.g.][]{eggleton2006EvolutionaryProcesses}. Under the stellivore hypothesis only binaries that are \emph{semi-detached} (i.e., not detached by a wide orbit, nor currently merging into a contact scenario), \emph{non-conservative} (i.e., they eject matter out of their gravitational well), and that are \emph{transient} (i.e. accretion is intermittent and not persistent) are candidates (see Fig.~\ref{fig:binarytypes}). As such, nova-likes (pink empty circles in Fig.~\ref{fig:controls}) are not good stellivore candidates because they display persistent accretion, and they are another control dataset.

Any astrophysics student learns that binary stars constitute a ``binary zoo'' \citep[for a popular-level introduction, see][]{lipunov1989WorldBinary}. The zoo metaphor illustrates that there is a large diversity of accreting binary stars, and even within the category that interests us, there are many more kinds of systems (see Table~\ref{tab:variety}).

\begin{table}[htbp]
\begingroup
\footnotesize
\centering
\begin{tabular}{@{}>{\raggedright\arraybackslash}p{0.36\linewidth} >{\raggedright\arraybackslash}p{0.32\linewidth} >{\raggedright\arraybackslash}p{0.28\linewidth}@{}}
\toprule
\textbf{White Dwarfs (WDs)} & \textbf{Neutron Stars (NSs)} & \textbf{Black Holes (BHs)}\\
\midrule
Cataclysmic variables & Accreting Millisecond X-Ray Pulsars (AMXPs) & Black Hole X-ray Binaries (BHXBs)\\[2pt]
\quad Nov\ae{} & X-ray bursters & Black hole transients (BHTs)\\[2pt]
\quad Dwarf nov\ae{} & Spider pulsars & Microquasars \\[2pt]
\qquad Z Cam stars & Transitional Millisecond Pulsars & \\[2pt]
\qquad SU UMa stars & & \\[2pt]
\qquad U Gem stars & & \\[2pt]
\quad Recurrent nov\ae{} & & \\[2pt]
\quad Nova-like variables & & \\[2pt]
\qquad VY Scl stars & & \\[2pt]
\qquad UX UMa stars & & \\[2pt]
\quad Magnetic Cataclysmic variables & & \\[2pt]
\qquad Polars & & \\[2pt]
\qquad Intermediate polars & & \\[2pt]
\qquad\quad DQ Her stars & & \\[2pt]
Symbiotic stars & & \\
\bottomrule
\end{tabular}
\caption{The variety of accreting binaries, classified according to the primary star (a dense WD, NS or BH). As typical entries to the literature, for binary stars in general see \citep{hilditch2001IntroductionClose,eggleton2006EvolutionaryProcesses}, for WDs, see \citep{warner1995CataclysmicVariable,hellier2001CataclysmicVariable}, and for accretion theory, covering NSs and BHs, see \citep{frank2002AccretionPower}.}
\label{tab:variety}
\endgroup
\end{table}

The phenomenology of binary stars is not only varied, it is also lively as it changes on human timescales (in milliseconds, seconds, minutes, hours, days, years or decades). This is remarkable because it is usually not the case in other astrophysical phenomena which vary on timescales of millions of years. This may explain why binaries in accretion have been and still are attracting the attention of amateur and professional astronomers and astrophysicists. As our observations of them continue, models of accreting binary stars have become ever more sophisticated, subtle and varied. This stellar diversity and richness is reminiscent of the biological realm.

Binaries are often hard to model (e.g. \citealp[pp.~254--256]{eggleton2006EvolutionaryProcesses} lists 18 general problems). The phenomenology of a particular system can be so unique that it is common to see papers or series of papers dedicated to a single particular binary star system.

Our point with these remarks about variety and difficulty of modeling is not to suggest any aliens-of-the-gaps arguments: ``if there is a phenomenon we cannot explain, then it's aliens'' \citep[see e.g. discussions in][]{wright2014InfraredSearch,vidal2019PulsarPositioning,vidal2026SearchTechnosignatures}. Obviously, one could list 200 anomalies or open problems, and we would still not have any proof of extraterrestrial life. More modestly, acknowledging enduring open issues, or enduring competing models to solve them does justify opening up the modeling space, also including astrobiological viewpoints. Once we have both astrobiological and astrophysical models, we can compare their predictive capabilities as well as their shortcomings. But we first need to be motivated to construct astrobiological models, and this is a core objective of this paper.

Note that, in a way, astrophysical models will never fail. Indeed, everything in the universe, whether it is living or not, can be described by physics. We could make a physics model of Albert Einstein by saying that he measured 172~cm. It would be a correct physical description, but we would be missing his biological, intellectual, political, and other human dimensions and subtleties.

Maybe astrophysics will continue its progress and explain all the phenomenology of accreting binary stars. Or maybe existing astrophysical models already contain ad hoc hypotheses and epicycles that would not be needed with astrobiological models. If astrobiological models become more predictive than astrophysical ones, we would be on the road towards a proof of extraterrestrial life.

\subsection{Homeostasis}

Another key feature of living and complex systems is \emph{homeostasis}, a capacity for remaining the same despite perturbations. \citet{derosnay1979MacroscopeNew} defined more precisely a homeostatic system as ``an open system that maintains its structure and functions by means of a multiplicity of dynamic equilibriums rigorously controlled by interdependent regulation mechanisms''. In other words, when perturbed, complex systems are able to come back to their own equilibrium, while simple physical systems tend to come back to thermodynamic equilibrium.

A known problem in the physics of WDs is that the boundary layer temperature is constant \citep[see][]{mukai2017XRayEmissions}, a phenomenon typical of homeostasis. For example, \citet{page2010SwiftObservations} found that the accreting WD V2491~Cyg stabilized its temperature after an initial rise, while \citet{zemko2015SuzakuObservation} found a similar temperature stable against orders of magnitude changes in luminosity.

\subsection{Geyser objection}

Besides the candle objection (see Sec. 2.2), a more sophisticated counter-example that involves a feedback loop is the \emph{geyser objection}. \citet[p.~118]{deacon2012IncompleteNature} describes the dynamics of the eruptions of the famous Old Faithful geyser at Yellowstone National Park in the following way:

\begin{quote}
``It erupts on a highly regular basis because of the way that the temperature and pressure of the subsurface water is self-regulated around a mean value by the interplay of geothermal heating, the pressure of steam, and the weight of the column of water in the underground vent. As the water is heated to the point of boiling, the pressure of the released steam reaches a value sufficient to drive the overlaying column of water upward; but in so doing it ``resets'' the whole system, allowing new water to accumulate, boil, and reach this same threshold, again and again. So the water temperature and steam pressure oscillate around a constant value and the geyser erupts at regular intervals''.
\end{quote}

Are accreting binaries not more sophisticated than the dynamics of a geyser? Basic astrophysical models of accreting binary stars indeed resemble the modeling of a geyser. One can point to two fundamental differences however. First, in a geyser there is no energy transformation; second, the eruptions of accreting binaries are not as predictable as those of geysers. The geyser system does show \emph{one single} natural feedback cycle. However, living systems have many feedback cycles, mediated by control mechanisms. Under the stellivore hypothesis, we expect more than a simple and trivial feedback cycle, i.e. intelligent control mechanisms.

High-energy interactions typically lead to destructive explosions if they are unchecked. For example, combining air and fuel leads to an explosion. However, if both are carefully put in a well-designed cylinder, they can form an internal combustion engine, a \emph{controlled explosion} that has proven revolutionary for our civilization. The same reasoning may apply to accretion and nuclear reactions. Unchecked nuclear reactions can and have led to catastrophes on Earth. In the universe, accretion or nuclear reactions can also rapidly lead to supernovae or other cosmic explosions. However, in the accreting binaries we have discussed, the energy is channeled via gravity and magnetic fields, and further nuclear reactions occur, sometimes slowly. The potential of nuclear reactions to organize complexity is also largely unexplored \citep[see however][]{maldonado-lang2025EmergentComplexity}.

\subsection{Fine-tuning}

A consequence of homeostatic control is that living or complex systems have to maintain their state within a narrow range of parameters to sustain their organization. If they deviate outside this range, they can die. This is why complex biological systems that are the products of a long evolutionary history appear to be delicately fine-tuned.

Similar situations occur in the modeling of accreting binaries. For example, the neutron star Aql~X-1 does not display pulsations, and yet has a strong magnetosphere. \citet[p.~7b]{patruno2018DeepPulse} conclude that ``any magnetosphere around this system requires a strong fine tuning of the parameters to explain the lack of pulses''. In a stellivore interpretation, one could speculate that modulation of the strength of the magnetosphere would be expected, as it is an essential variable that can control others (e.g. the accretion rate).

Another example is in the formation of millisecond pulsars (MSPs). \citet{smedley2017ImplicationsCompanion} have simulated the standard formation model and found that ``the range of initial orbital periods required to produce the currently observed range of orbital periods of MSPs is extremely narrow''. They do propose an alternative model that mitigates the issue, but it is beyond the scope of this paper to evaluate it. In a stellivore interpretation, the standard evolutionary formation model may indeed be wrong, and other paths to form binaries involving some kind of stellar engineering need to be explored.

Here, in particular, MSPs play the role of a timing and positioning standard for the galaxy, called the Pulsar Positioning System \citep[see][]{vidal2019PulsarPositioning}. If the Pulsar Positioning System were engineered, the formation of MSPs would indeed have a different history. For example, active MSPs would require serious cooling to function properly, and helium is an excellent coolant choice. This would be an alternative explanation for why most MSPs are found accreting a helium WD.

\subsection{Reserves}

Biology produces robust systems that can absorb unknown perturbations and keep their metabolism working thanks to reserves \citep[see e.g.][]{kooijman2009DynamicEnergy}. The crucial role of reserves is also well-known in cybernetics and engineering. In the dynamics of accretion and ejection displayed by binaries, there is most often an accretion disk that temporarily stores plasma, except for polar cataclysmic variables.

There is another less obvious buffering that exists through the observation of accreting NSs, namely that the ``burst rate decreases for most sources with increasing \Mdot{} [accretion rate], despite the more rapid accumulation of fuel'' \citep[p.~397a]{galloway2008ThermonuclearType}. This observation is indeed counterintuitive from a purely physical perspective, as one would expect bursts expelling matter out of the system to increase, and not decrease, as the accretion rate rises. This means that the fuel has to be stored, possibly in a controlled manner.

There is another independent argument showing that such storage happens, namely the fact that nuclear physics plays an important role in the phenomenology of NSs, while the gravitational energy released by accretion is about 40 times more efficient than nuclear fusion and should thus make nuclear reactions almost irrelevant. As \citet[p.~113]{strohmayer2006NewViews} put it:

\begin{quote}
if the accreted fuel was burned at the rate of accretion, any evidence of nuclear physics would be swamped by the light from released gravitational energy. The only way the nuclear energy can be seen is when the fuel is stored for a long period and then burns rapidly (as in Type I bursts and superbursts).
\end{quote}

A third indication that there are reserves is that in many sources, the bursts are not frequent enough to burn all the accreted fuel \citep{vanparadijs1988RelationXRay,zand2003Superburst4U}. This has led to suggestions that another kind of slower nuclear burning is happening \citep[see e.g.][]{bildsten1995PropagationNuclear,revnivtsev2001NewClass}. In the stellivore interpretation, this might be a sign that some energy is also used internally for growth and maintenance or that controlled nuclear reactions are taking place.

\subsection{Accretion control}

A high energy flow or high power density is not a sufficient condition for life (see section 2.2). 
The \textit{control} of energy flow is equally essential for life. 
For example, if a human ate non-stop or stopped eating completely, they would end up dying.
So energy flow needs to be controlled to ensure the basic functions of growth, maintenance and reproduction \citep[see e.g.][]{aunger2007RigorousPeriodization,chaisson2011EnergyRate}. 
In the stellivore context, the challenge is thus to find signs of \emph{accretion control}. On a superficial level, it is already well known and well studied that accretion turns itself on and off in many binaries. More precisely, there are really two kinds of accreting NSs or BHs: the \emph{persistent} ones that accrete at a high rate and constitute about two thirds of known systems, and the remaining third are \emph{transient} ones with lower accretion rates \citep{tanaka1996XRayNovae}. The transient ones are thus the only stellivore candidates because of their observed accretion pattern switching on and off. It is difficult to completely rule out the persistent ones as they may be persistent only for the time scale we have been observing them.

Let us take a closer look at how accretion turns itself on and off. In the case of magnetic WDs, it would make the most sense to control accretion through magnetic field lines. For example, the accretion stream in AM~Her stars typically displays a complex ``transition from ballistic to magnetically controlled flow'' \citep[pp.~111--112]{hellier2001CataclysmicVariable}. One of the reasons for this complex transition is that the magnetic field is tilted, so that the plasma stream preferentially follows this route. 
Hellier remarks that physical systems ``tend to settle into their lowest-energy configuration, but diverting the stream out of the plane requires energy''. We again have a far-from-equilibrium situation. As we suggested earlier \citep[pp.~240--241]{vidal2014BeginningEnd}, in the stellivore interpretation, tilting the magnetic field would be a straightforward way to turn accretion on and off.

In the case of NSs, the accretion rate can vary over 4 orders of magnitude \citep[p.~93]{lamb2008AccretingNeutron}. The accretion rate can be modulated via the accretion disk, or via a variation of the magnetic field's strength, which itself can vary over 4 orders of magnitude. For accretion control via an accretion disk, \citet{bult2018NICERLook} suggest accounting for recent observations of Aql~X-1 by ``an intrinsically variable accretion disk that propagates variability down to the power-law emitting region by modulating the mass accretion rate''.

More generally ``there is strong evidence that the external magnetic fields of neutron stars in LMXBs [Low Mass X-ray binaries] decrease by factors $\sim 10^2$ - $10^3$ during their accretion phase perhaps on timescales as short as hundreds of years'' \citep[p.~94]{lamb2008AccretingNeutron}. However, the magnetic field can also change on much shorter timescales, for example the system SAX~J1748.9--2021 ``was observed first as a non-pulsating atoll source (in 1998), then it turned into an intermittent AMXP [Accreting Millisecond X-ray Pulsar] (in 2001, 2005 and 2009) [...], and then it became a persistent AMXP in 2015 [...]. In this case therefore, the neutron star magnetosphere, if absent in 1998, must have re-emerged on a relatively short timescale for a reason that is not completely clear'' \citep{patruno2018DeepPulse}.

To sum up, in the stellivore interpretation, lowering or tilting the magnetic field would allow accretion control. In the case of BHs, the dynamics is similar to that of NSs, but the control might be interpreted via the different states of the BH accretion disk, which are known to correspond to different levels of accretion rate.

\subsection{Ejection Control}

For a living being or a society to prosper in the long run, disposing of waste is as critical as ingesting new food or energy. So one should also look at traces of controlled ejection of waste materials in putative stellivores. Generally, WDs eject matter through \emph{novae}, while NSs and BHs eject via \emph{bursts} and \emph{jets}.

In the case of accreting WDs, astrophysicists have found a ``growing sample of novae which show evidence for complex, multi-phase ejection'' \citep{chomiuk20142011Outburst}.

In NSs, double thermonuclear bursts (Type I) have been found. As \citet{galloway2008ThermonuclearType} comment: ``bursts with extremely short recurrence times (`double' or `prompt' bursts) have long presented a challenge to our understanding of burst physics. Their recurrence times of $\gtrsim 5$ minutes are too short for sufficient fuel to accumulate to allow ignition by unstable thermonuclear burning''. As \citet[p.~254]{lewin1993XrayBursts} comment, ``these intervals are much too short to replenish, through accretion, a sufficient amount of nuclear fuel to account for the second burst. Thus, one requires the presence of a reservoir of nuclear fuel which survived the previous thermonuclear flash, and can be prematurely rekindled''. In the stellivore interpretation, this is consistent with the existence of reserves, and with the ability to use just a part of them.

Even more intriguingly, in the three NS systems 4U~1636-536, 4U~1709-267 and SAX~J1808.4-3658, it has been found that a weak burst \emph{precedes} a normal burst by only a few seconds  \citep{galloway2008ThermonuclearType}. A similar behavior has also been observed with the WD system T~CrB, about which \citet{schaefer2010ComprehensivePhotometric} comments: ``how can the turning off of accretion \emph{anticipate} or \emph{trigger} the nova event?''. It is not clear how to interpret such behavior, but anticipation is indeed a key hallmark of life, and we might be dealing with ejection control here. Finally, \citet{schaefer2026OrbitalPeriod} has shown that the full eruption cycle (from accretion to ejection) is actually insufficient to lead to a supernova. This is consistent with a living interpretation where the system preserves its own survival and avoids its destruction.

\subsection{Living system summary}

To sum up the living cues we have reviewed, the framework of living systems theory \citep{miller1978LivingSystems,miller1990Timer} is most useful \citetext{\citealp[pp.~239--248]{vidal2014BeginningEnd}; \citealp{vidal2016StellivoreExtraterrestrials}}. As summarized in Table~\ref{tab:subsystems}, J.~G.~Miller argues that all living systems have 20 subsystems performing different critical functions (the 20$^{\mathrm{th}}$ timer subsystem was added later by J. L. Miller).

\begin{table}[htbp]
\footnotesize
\centering
\begin{tabular}{lll}
\toprule
\makecell{\textbf{Matter + Energy } \\ \textbf{+ Information}} & \textbf{Matter + Energy} & \textbf{Information}\\
\midrule
1. Reproducer & 3. Ingestor & 11. Input transducer\\
2. Boundary & 4. Distributor & 12. Internal transducer\\
 & 5. Converter & 13. Channel and net\\
 & 6. Producer & 14. Decoder\\
 & 7. Matter-energy storage & 15. Associator\\
 & 8. Extruder & 16. Memory\\
 & 9. Motor & 17. Decider\\
 & 10. Supporter & 18. Encoder\\
 & & 19. Output transducer\\
 & & 20. Timer\\
\bottomrule
\end{tabular}
\caption{Twenty living subsystems proposed by J.~G. and J.~L.~Miller. They are classified according to their use of matter, energy and information (left column), matter and energy (central column), or information (right column).}
\label{tab:subsystems}
\end{table}

In the stellivore interpretation, the \emph{reproducer} subsystem is the most speculative one, yet arguably the most fundamental. Stellivores would be tackling the ultimate issue any intelligence in the universe has to face: to find a solution to the heat death of the universe. One way out would be to make a new offspring universe through a bounce in the singularity of a black hole, which would certainly require enormous amounts of energy (see \citealp{vidal2014BeginningEnd} for a book-length case of this scenario and references therein). In this light, the high power density value of SMBHs would be an expected feature in an evolutionary cosmology where black holes play a key role \citep[see e.g.][]{smolin1992DidUniverse}, even if no intelligent life is involved.

In this paper we have focused on matter-energy subsystems, and it is already clear that the accreting binaries we have considered have \emph{boundaries} (2), \emph{ingestor} functions through accretion (3), \emph{matter-energy storage} through reserves (7), and an \emph{extruder} function (8) through ejecta and jets.

The \emph{distributor} function (4) is less obvious because distribution around the components of an organism happens at a lower scale, but the stellivore interpretation would indeed expect mechanisms to distribute the energy on or near the surface of WDs, NSs and BHs.

The \emph{converter} (5) function can be inferred indirectly, by remarking that energy is transformed between what is accreted and what is ejected. For example in accreting WDs, the composition of nova ejecta displays unusual heavy-element abundance, and such heavy elements are not present in the material accreted from the companion star \citep{prialnik2001Novae}. The hypothesis that heavy elements are produced during the nova has to be ruled out, because the temperature is not high enough to produce elements heavier than helium. There remain two possibilities. Either the accreted matter is somehow mixed with WD material (mainstream astrophysical picture), or the accreted material is used as fuel to perform work and produces waste as heavy elements (stellivore picture).

Regarding the \emph{motor} function (9), there are binaries that are moving fast in the galaxy (e.g. the extremely low-mass WD J0755~+~4906, the neutron star IGR~J11014~-~6103 or the black hole XTE~J1118~+~480). \citet[p.~260]{vidal2014BeginningEnd} predicted that if stellivores are alive, the motion of higher-velocity binaries should not be random, but directed toward a nearby star, because they would be looking for the nearest next food source. We could also predict that higher-velocity binaries have on average lower-mass companions, meaning that their energy source is almost exhausted, and they need to find and accrete a new star. Such predictions of energy-seeking behavior are consistent with observations of spider pulsars, suggesting further testable intelligent behavior \citep{vidal2024SpiderStellar,vidal2023AreSpider,vidal2026SpiderStellar}.

The \emph{supporter} function (10) maintains the proper spatial relationships between the components. One could argue that the orbital and magnetospheric parameters contribute to this support function if they are indeed fine-tuned to ensure the living functions, and if they keep adapting for efficient controlled accretion, controlled ejection, and controlled motion.

What is crucially missing in this picture are the information subsystems (third column in table~\ref{tab:subsystems}). The notable exception is the \emph{timing} (20) subsystem, with the Pulsar Positioning System that exists in our galaxy thanks to MSPs accelerated through accretion, a galactic navigation system intriguingly and remarkably accurate down to 100 meters. \citet{vidal2019PulsarPositioning} made a detailed analysis of how to test whether MSPs might have been engineered, and argued that even if the engineering hypothesis is wrong, it remains extremely useful for any civilization in our Milky Way that would need to set galactic timing, navigation and metadata communication standards \citep{vidal2017MillisecondPulsars}.

It is also worth noting that the traditional search for extraterrestrial intelligence (SETI) looks only at information signals in the environment, the \emph{output transducer} (19), and has found none so far. By contrast, by broadening the search to all 20 living subsystems, the stellivore interpretation suggests that the behavior of up to 10 of them is consistent with, or has already been observed in accreting binaries.

Additionally, one might be seeing a hierarchy of power density, which is higher for more compact accreting objects, from WDs, to NSs and BHs. In the stellivore interpretation, following the Barrow scale of civilizational development, the most compact would be the most advanced too \citep[sec.~9.2.2]{barrow1998ImpossibilityLimits,smart2012TranscensionHypothesis,vidal2014BeginningEnd,vidal2016StellivoreExtraterrestrials,vidal2026SearchTechnosignatures}.

We would also like to emphasize that we are looking at macroscopic features of accreting binaries, and their power densities may be better compared with Earth seen from space, which we saw was about $2.9 \times 10^{-8}$~\Wkg.

\section*{Conclusion}

We have proposed a new universal signature for astrobiology: to simply follow the power density. 
Rather than an unambiguous signature, we see high power density as a necessary but not sufficient condition for assessing whether a system is living or complex. 
It could also be applied to any candidate living system at different scales, such as monitoring the power density of in situ missions in the Solar System.

In this paper, we focused on a very large scale by computing the power density of many binary star systems in accretion and revealed that their power density is high. We reviewed a control sample of accreting young stellar objects, showing that their power density is clearly below that of Earth's living systems and candidate stellivores. However, two other control samples, nova-like cataclysmic variables and supermassive black holes, also have high power density values. This motivated us to look for other living clues regarding accreting binaries, such as their variety, their existence far from thermodynamic equilibrium, their apparent homeostasis, the need to fine-tune models to explain them, the apparent existence of energy reserves, accretion control and ejection control, as well as other potential living systems functions such as motor function or timing.

Power density remains a useful complexity metric, and we saw the need to carefully thread through issues of defining a system's boundary and assumptions about energy efficiency. Big history and astrobiology could also use new informational or computational metrics in addition to or instead of power density \citep[see e.g.][]{delahaye2018OrganizedComplexity,sowinski2025CausalLeverage}. If the stellivore hypothesis is true, then complexity would have risen dramatically at other places in the galaxy, and the story of rising complexity would not be centered locally on planet Earth anymore, but would be much bigger, richer, and more fascinating.

\section*{Data availability}

Our datasets and pipeline are available at:\\
\url{https://github.com/luchprojects/Power-Density-Biosignature-Figures}

\section*{Acknowledgments}

Cl\'ement Vidal presented an early version of this paper at the International Big History Association (IBHA) \& INAF-IASF MI Symposium, 15-16 July 2019, ``Life in the Universe: Big History, SETI and the Future of Humankind'', Milan, Italy. We used Cursor to perform and double-check some calculations.

\printbibliography

\end{document}